\input{aipcheck}

\documentclass[
    ,final            
  ]
  {aipproc}

\layoutstyle{6x9}

\begin{document}
\newcommand{\tc}{\rm T_c}
\newcommand{\jpsi}{J / \psi}
\newcommand{\ec}{\eta_c}
\newcommand{\asx}{\chi_{c_i}}
\newcommand{\gt}{G(\tau, \vec{p}, T)}
\newcommand{\gtb}{G(\tau, T)}
\newcommand{\greconb}{G_{{\rm recon}, T^*}(\tau,T)}
\newcommand{\sgw}{\sigma(\omega, \vec{p}, T)}
\newcommand{\sgb}{\sigma(\omega,T)}
\newcommand{\cb}{\chi_b}
\newcommand{\ub}{\Upsilon_b}
\newcommand{\eb}{\eta_b}

\title{Quarkonia in a deconfined gluonic plasma}

\classification{11.15.Ha, 12.38.Gc, 12.38.Mh, 25.75.-q}
\keywords      {Quark-gluon plasma,quarkonia suppression}

\author{S. Datta}{
  address={Brookhaven National Laboratory, Upton, NY 11973, USA}
}

\author{A. Jakov\'ac}{
  address={Institute of Physics, BME Budapest, H-1111 Budapest,
  Hungary}
}

\author{F. Karsch}{
  address={Brookhaven National Laboratory, Upton, NY 11973, USA},
  altaddress={Bielefeld University, D-33615 Bielefeld, Germany}
}

\author{P. Petreczky}{
  address={Brookhaven National Laboratory, Upton, NY 11973, USA}
}

\begin{abstract}
We discuss lattice results on the properties of finite momentum
charmonium states in a gluonic plasma. We also present preliminary
results for bottomonium correlators and spectral functions in the 
plasma. Significant modifications of $\chi_{b_{0,1}}$ states are seen
at temperatures of 1.5 $\tc$. 
\end{abstract}

\maketitle


Following the suggestion of Matsui and Satz \cite{satz} that 
$\jpsi$ can act as a probe of deconfinement, heavy quarkonia in 
the context of relativistic heavy ion collisions have been extensively
studied both theoretically and experimentally. Early, potential
model based studies indicated that all charmonium bound states dissolve
by temperatures $\sim$ 1.1 $\tc$ \cite{karsch1}. But direct lattice
studies over last few years have concluded that while the excited 
$\chi_c$ states dissolve quite early in the plasma \cite{prd}, the 
1S states $\jpsi, \ec$ survive till quite high temperatures 
\cite{prd,others}, at least in a purely gluonic plasma 
\footnote{Naively, one would not expect a much earlier dissolution 
due to dynamical quarks. A recent study in 2-flavor QCD 
\cite{morrin} supports this expectation.}. 
It has recently been claimed
\cite{karsch2} that the observed $\jpsi$ suppression in SPS
and in RHIC is consistent with suppression of only the
secondary $\jpsi$ from excited state decays, in accordance with the
lattice results.

For understanding the mechanism of charmonia dissolution, as well as
for phenomenological purposes, it is important also to know the effect of
the plasma on a bound state in motion with respect to the plasma
rest frame. This problem can be studied directly on lattice, in ways 
similar to that of the bound states at rest \cite{sewm}. 
We look at the momentum-projected Matsubara correlators
\begin{equation}
\gt = \sum_{\vec{x}} e^{i \vec{p}.\vec{x}} \; \langle J_H(\tau, \vec{x})
J_H^\dagger (0, \vec{0}) \rangle_T
\end{equation}
where $J_H$ is a suitable mesonic operator, $\vec{p}$ the spatial 
momentum, $T$ is the temperature of the gluonic plasma and the 
Euclidean time $\tau \in [0,1/T)$.
Through analytic continuation, the Matsubara correlator can be 
related to the hadronic spectral function by an integral equation:
\begin{equation}
\gt =\int_0^{\infty} d \omega
\; \sgw \; \frac{\cosh(\omega(\tau-1/2
T))}{\sinh(\omega/2 T)}.
\label{eq.spect}
\end{equation}

The 1S charmonia $\ec, \jpsi$ at rest undergo very little significant 
modification till temperatures of 1.5 $\tc$. However, the finite 
momentum correlators $\gt$ show significant temperature dependence even
earlier. Medium modifications become stronger with
increasing momentum. For $p \sim$ 1 GeV, significant modifications of
the correlator are already seen at 1.1 $\tc$. The spectral function
$\sgw$, extracted from $\gt$, shows a clear peak also at high
momenta, but it is significantly modified from the zero temperature
peak. Physically this can be understood as follows. A charmonium state
moving in the plasma frame ``sees'' more energetic gluons, leading to
an increase in its collisional width. An in-medium change of
the energy-momentum dispersion relation is also possible. 
Due to paucity of space, we refer 
to Ref. \cite{sewm} for further discussion of this, and turn 
here to bottomonia instead. 

The $\ub$ peak in the dilepton channel will be accessible to both RHIC and 
LHC, and may produce cleaner setups for plasma-related modifications since
normal nuclear modifications for bottomonia are expected to be
small. On the other hand, the 1S states $\ub$ and $\eb$ are very tightly
bound and even the potential model calculations estimated a very high 
dissolution temperature for them \cite{karsch1}. The behavior of 1P
bottomonia is less clear: while the potential models predict a dissolution 
temperature close to $\tc$ \cite{karsch1}, they 
also suggest a size similar to the 1S charmonia
for these states and therefore one may expect a similar 
dissolution temperature \cite{karsch1}. A recent study \cite{bottom},
on the other hand, has found modifications of 
$\cb$ close to $\tc$, unlike $\eb$. More than 40\% of the 
total $\ub$ seen in hadronic
collisions come from decay of excited bottomonia, and an early 
dissolution or strong modification of $\chi_b$ will modify this 
contribution significantly. 

\begin{center}
\begin{figure}[h]
 \includegraphics[width=7.5cm]{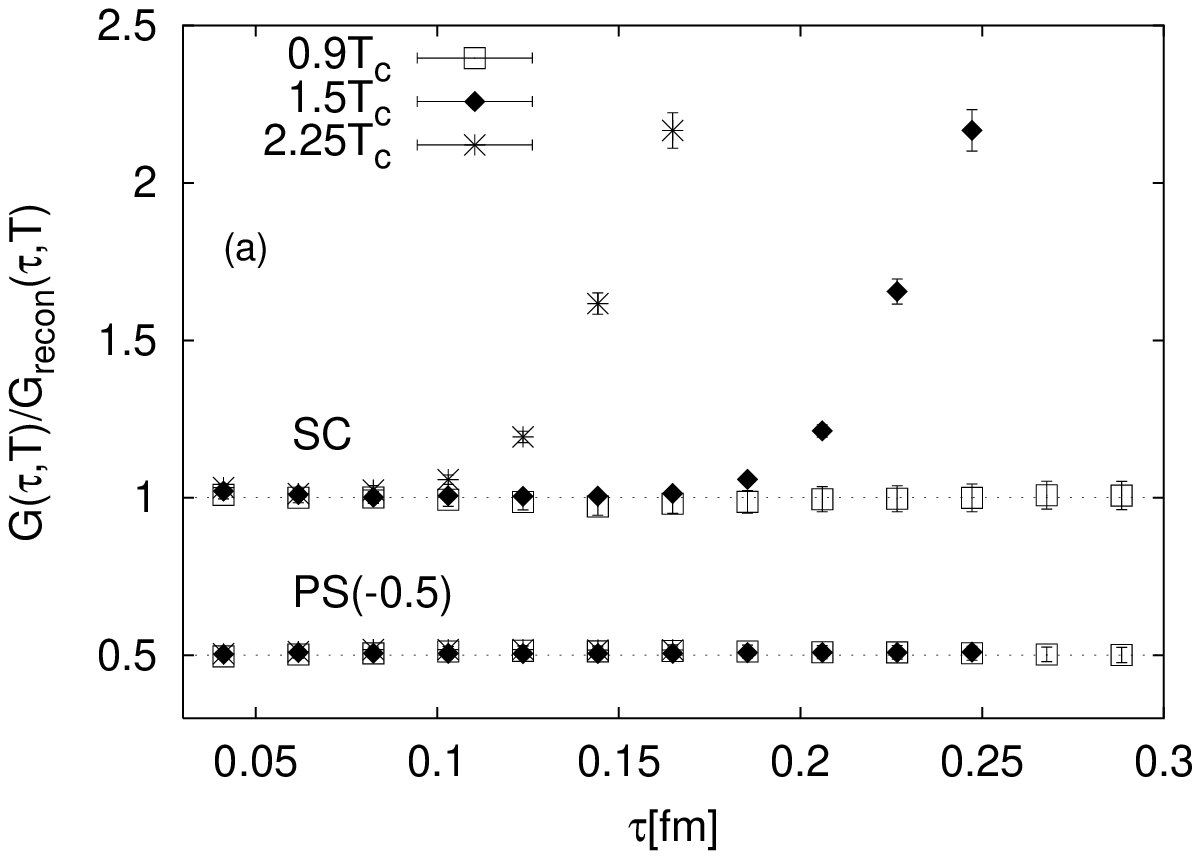} 
\includegraphics[width=7.5cm]{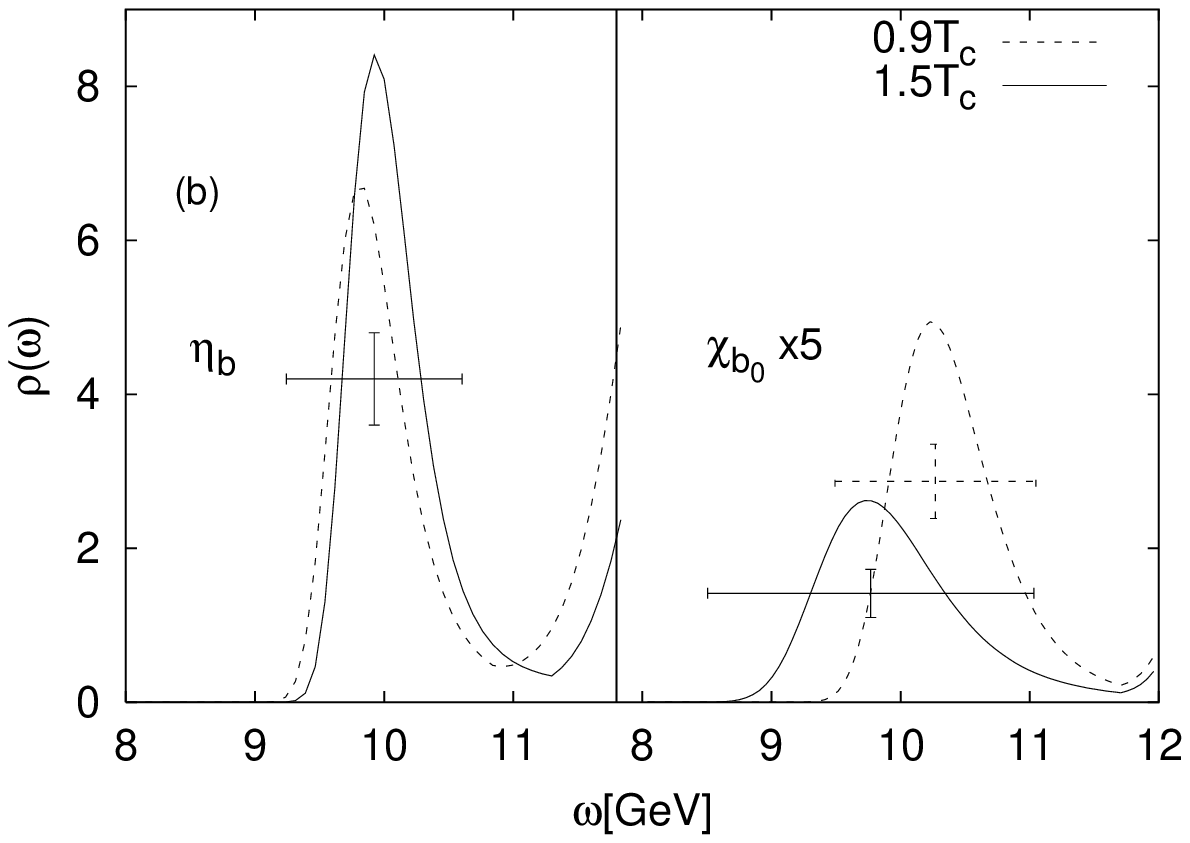}
  \caption{(a) $G(\tau,T)/G_{\rm recon}(\tau,T)$ for $\bar{b} \gamma_5
  b$ and $\bar{b} b$ at $\vec{p}=0$. (b) Spectral function constructed from 
$G(\tau,T)$ using maximum entropy method.}
\label{fig.bottom}
\end{figure}
\end{center}

We studied bottomonia in gluonic medium following the same methods
used for the charmonia study in Ref. \cite{prd}, and on the finest set
of lattices used there. These lattices have a cutoff $a^{-1}$ = 9.72
GeV, which is somewhat coarse for bottomonia
\footnote{Ref. \cite{bottom}, uses an anisotropic 
lattice which is slightly finer in time, $a_t^{-1}$ = 
10.89 GeV, but considerably coarser in space, $a_s^{-1}$ = 2.72 GeV. It
also uses a different action, so the cutoff effects should be different.}.
This, therefore, should only be taken as a pilot study. We studied zero 
momentum projected $\bar{b} \Gamma b$ (point-point) correlators, where
$\Gamma = \gamma_5, \gamma_i, 1$ and $\gamma_i \gamma_5$ for $\eb, \ub,
\chi_{b_0}$ and $\chi_{b_1}$, respectively. 

We extract $\sgb$ from $\gtb$ using the
``Maximum Entropy Method'' \cite{mem}, 
where the inversion of Eq. (\ref{eq.spect}) is turned into a
well-defined problem of finding the most probable spectral function
given data and prior information for $\sgb$. Also very useful and robust
conclusions of possible change of state with deconfinement can be
obtained by comparing the correlators measured above $\tc$ with
$\greconb$, correlators reconstructed from the spectral function 
obtained at the smallest temperature below $\tc$ (see Ref. \cite{prd}
for details of our analysis method). If the spectral function is
not modified with temperature, $\gtb / \greconb$ = 1.
A comparison of the measured 
correlators in the pseudoscalar and scalar channels with the
reconstructed correlators is shown in Fig. \ref{fig.bottom}(a).
$\gtb$ for the pseudoscalar shows no significant modification 
for temperatures upto 2.25 $\tc$, indicating that $\eb$ is essentially
unmodified at these temperatures. The scalar correlator, on the other
hand, shows large changes at long distances already at 1.5 $\tc$.
The modification pattern in  Fig. \ref{fig.bottom}(a) is somewhat
different from that seen in scalar charmonia, where the medium effect 
was seen to have set in at smaller distances and less abruptly.

Figure \ref{fig.bottom}(a) shows qualitatively similar trend as in 
Ref. \cite{bottom}, but the deviation of the
scalar correlator seen by us is much smaller than that seen in
Ref.\cite{bottom}. This could be due to the use of anisotropic lattice 
in \cite{bottom}. Figure \ref{fig.bottom}(b) shows a comparison of 
the ground state peaks at 0.9 $\tc$ and 1.5 $\tc$ for the pseudoscalar (left)
and scalar (right) channels. Plotted here is the dimensionless
quantity $\rho(\omega,T) = \sigma(\omega,T) / \omega^2$.
As expected, the $\eb$ peak shows
no significant modification at 1.5 $\tc$. The $\chi_{b_0}$ peak, on the
other hand, shows significant deviation, with a possible shift and
broadening. The $\ub$ and $\chi_{b_1}$ show similar trends
to $\eb$ and $\chi_{b_0}$, respectively. It will be interesting to
further study the modification of the $\chi_b$ peak, both in terms of
the nature of the modification and its behavior at temperatures 
closer to $\tc$.


This manuscript has been authored under contract number 
DE-AC02-98CH1-886 with the US Department of Energy. 
Computations were done in NERSC, Berkeley and in NIC, Juelich.
A.J. is supported by Hungarian Science Fund OTKA (F043465).


\end{document}